# Why does Boltzmann's ergodic hypothesis work and when does it fail

M. Howard Lee[a,b,*]

[a]*Physics Department, University of Georgia, Athens, GA 30602 USA\* and*
[b]*Korea Institute for Advanced Study, Seoul 130-012, Korea*

[*]Permanent address

*E-mail address:* mhlee@uga.edu (M.H. Lee).

**Abstract**

According to a recently given ergodic condition for Hermitian many-body models the thermodynamic limit and irreversibility are necessary but by themselves not sufficient. The sufficient condition turns out to be the existence of a zero frequency mode. It is measured by an infinite product of the recurrants from the recurrence relations method, which solves the Heisenberg equation of motion in Hermitian models. This condition has been tested with a variety of assemblies of nearest-neighbor coupled harmonic oscillators. The results provide a physical insight into why the ergodic hypothesis is valid and when it fails.

## I. Introduction

Boltzmann's ergodic hypothesis is widely accepted in statistical mechanics [1]. But there exists evidence that it may not be universally valid [2]. Thus it is desirable to know the physics underlying the hypothesis. Ergodic theory is a branch of mathematics originally founded to establish the validity of the hypothesis. But to date it has shed little light on why some physical models should be ergodic while some others not ergodic.

To address this issue more physically, we recently have taken the time average of a time dependent susceptibility and compared it with its time independent counterpart [3]. Therewith we have obtained a necessary and sufficient condition for ergodicity. Since the dynamic susceptibility is related to the inelastic scattering processes, we can attribute ergodicity to how the energy imparted by an external probe becomes delocalized in a system or a scatterer. In this work we shall illustrate the connection between ergodicity and energy delocalization via several different assemblies of nearest-neighbor (*nn*) coupled harmonic oscillators (*HOs*). A similar analysis but using electron gas models is given elsewhere [4].

## II. Ergodic condition

We shall very briefly recall the ergodic condition previously obtained. Let $H$ be the Hamiltonian describing a system of interest, assumed to be *Hermitian*, and $A$ a dynamical variable. We can construct the relaxation function $r(t) = (A(t), A)/(A, A)$, where $A(t) = \exp itH \, A \exp -itH$, $A = A(t=0)$, $\hbar = 1$. The necessary and sufficient condition for ergodic behavior of a model with respect to dynamical variable $A$ is that $W$ be finite, i.e., $0 < W < \infty$, where

$$W = \int_0^\infty r(t) \, dt \tag{1a}$$

$$= \tilde{r}(z=0), \tag{1b}$$

where $\tilde{r}(z)$ is the Laplace transform of $r(t)$. Equivalently it may also be expressed in an infinite product studied by Wallis, to be referred to as the *canonical* form,

$$W = \frac{\Delta_2 \cdot \Delta_4 \cdot \Delta_6 \cdot \mathrm{K}}{\Delta_1 \cdot \Delta_3 \cdot \Delta_5 \cdot \mathrm{K}} \tag{1c}$$

where $\Delta$'s are recurrants defined in the recurrence relations method. Although the three ways of calculating $W$ are equivalent, in practice one way may be simpler than the others as we shall see in our examples in Section III.

## III. Ergodic condition and harmonic oscillator models

Although our formulation is constructed for quantum models, it is equally applicable to classical models. In this work we choose models of classical nn coupled *HOs*, for they render simple interpretation for the ergodic condition. The Hamiltonian is the usual one,

$$H = \sum_i^N p_i^2/2m_i + k/2 \sum_{(ij)}^N (r_i - r_j)^2 \tag{2}$$

where all the symbols have the standard meaning. There is a rich variety in the way this model can assume by our choice on mass $m_i$ and the lattice structure.

*A. Monatomic chains:* $m_i = m$ and $(ij) = (i \ i+1)$.

*A.1. Monatonic chain in periodic boundary conditions*

We choose $A = p_0$, the momentum of any one of $N$ atoms in the chain, which has been perturbed by a probe. We assume $N \to \infty$. The relaxation function $r(t)$, also known as the momentum autocorrelation function of $p_0$, in this familiar *HO* chain is well known: in units of $k/m = 1$,

$r(t) = J_0(2t)$ and $\tilde{r}(z) = (z^2 + 4)^{-1/2}$, where $J_0$ is the Bessel function [5]. Using these results in (1a and 1b)) we can now at once obtain $W = 1/2$. Evidently it is also corroborated by the canonical form (1c)

$$W = \frac{1 \cdot 1 \cdot 1 \cdot 1 \cdot K}{2 \cdot 1 \cdot 1 \cdot 1 \cdot K} = \frac{1}{2}. \quad (3)$$

Since $W$ is finite, with respect to $p_0$, the *HO* chain is ergodic as long as $N \to \infty$. At the thermodynamic limit, the autocorrelation function is irreversible. That is, the perturbation energy imparted at $0^{th}$ mass becomes delocalized, first to its *nn*'s and then down to the next. As $t$ grows large, it never returns to the original atom if $N \to \infty$ first. All the atoms are sampled. More precisely said, there exists a zero frequency mode, indicated by a coherent translation mode involving all the masses in the chain. If $N$ is finite, there is a recurrence and $r(t)$ is periodic. Such a system is of course not ergodic since it cannot admit irreversibility, a necessary condition for ergodicity.

Also observe that if $A = p_0$ as in this case, $\tilde{r}(0)$ corresponds to the diffusion constant according to linear response theory [6]. There is another model that has similar dynamical features. It is a 1*d* system of hard rods of point mass in an infinitely long ring [7]. With respect to $A = p_0$, the model is also ergodic. An initial collision sets in a chain of collisions involving ultimately all the particles in the chain. There is a finite diffusion constant.

### A.2. Monatomic chain between walls

Now we let $m_0$ and $m_{N+1}$ be infinitely heavy and $m_i = m$ if $i = 1, 2, K\ N$, so that it is a *HO* chain attached to a wall on each end. For *A* we could choose the momentum of any *HO* at a finite distance from say the left wall (the position of the $0^{th}$ oscillator with infinite mass). The simplest is to choose $A = p_1$. For this system it is known that if $N \to \infty, r(t) = J_0(2t) - J_4(2t)$, again in units of $k/m = 1$ [5b]. The canonical form is as follows:

$$W = \frac{1/2 \cdot 2/3 \cdot 3/4 \cdot K}{2/1 \cdot 3/2 \cdot 4/3 \cdot K} \quad (4)$$

It would seem that $W = 0$. Indeed $\int [J_0(2t) - J_4(2t)] dt = 0$, indicating that this system is not ergodic with respect to $p_1$, or with respect to the momentum of any other oscillator at a finite distance from the left wall.

The underlying physics is revealing. When the *HO* at site 1 is perturbed, it sets up a standing wave. The atoms at the nodal positions are stationary and do not participate in oscillation. The perturbed energy is thus not fully delocalized and hence not ergodic. Vanishing of $\tilde{r}(0)$ of course means that there is no diffusion. A chain that is "nailed down" cannot have translation.

### A.3. HOs on a Bethe Lattice

Let $m_i = m$ for all $i$, but $(ij)$ is such as to form a Bethe lattice with $q$ nn's. If $q = 2$, the lattice reduces to a $1d$ chain of A.1. If $q = 3$, it is a disjointed lattice generated by bifurcation. It is not a regular Bravais lattice, a quasi lattice, but not without interest owing to a self-similarity property. For $A$, we may choose the momentum of any oscillator since it is all equivalent. If $N \to \infty$, the canonical form is as follows in the units $k/m = 1$ [8]: For $q \geq 2$,

$$W = \frac{1 \cdot 1 \cdot 1 \cdot K}{2 \cdot (q-1) \cdot (q-1) \cdot K} \tag{5}$$

Observe that if $q = 2$, it recovers that for a monatomic chain A.1. It is evident that $W = 0$ if $q \geq 3$, as borne out by $\tilde{r}(z \to 0) = az + 0(z^2)$, where $a$ is a constant.

There is no coherent translation in the Bethe lattice, that is, there is no diffusion in the system. When an oscillator is perturbed, evidently the perturbation energy does not become delocalized to all other oscillators owing to a quasi lattice structure. An excitation in one branch may not reach oscillators in another branch if $N \to \infty$. There is thus no coherent translation mode, meaning no zero frequency model.

### B. Independent Oscillator (IO) model

Now $m_0 = M$, $m_i = m$, for $i = 1, 2, \ldots N$, with $mN/M = \lambda < \infty$; and $(ij) = (i0)$ only. It is a model for Brownian motion, apparently due to Zwanzig. Observe that it is still a Hermitian model, hence satisfying Hermitian dynamics, not stochastic dynamics as do most models for Brownian motion.

We choose $A = p_0$, the momentum of the Brownian or $B$ particle, to which small oscillators are coupled harmonically. The canonical form in units of $k/m = 1$ is as follows [9]:

$$W = \frac{(3 \cdot 3/3 \cdot 5)(5 \cdot 5/7 \cdot 9)(7 \cdot 7/11 \cdot 13)K}{(\lambda/1 \cdot 3)(2 \cdot 2/5 \cdot 7)(4 \cdot 4/9 \cdot 11)K} = \infty \tag{6}$$

The above result is borne out by $\tilde{r}(z) = const/z$ as $z \to 0$. There is no zero frequency mode. When the $B$ particle is perturbed, it suffers a translational motion. Those small particles attached to it are dragged or pulled along and they do not necessarily follow coherently. Thus, the behavior is not ergodic.

### C. Monatomic chain with one impurity

The model is the same as described in A.1 except that one particle has a different mass $m_0$. We define $\lambda = m/m_0$. If $\lambda$ is large, it is a light mass impurity, eventually reaching a vacancy. If $\lambda$ is small, it is a heavy mass impurity, which can act like a Brownian particle. In that limit it is another model of Brownian motion, comparable to the $IO$ model given in B.

We choose $A = p_0$. For this dynamical variable, in units of $k/m = 1$ the canonical form of $W$ is [5a]

$$W = \frac{1 \cdot 1 \cdot 1 \cdot 1 \cdot K}{2\lambda \cdot 1 \cdot 1 \cdot 1 \cdot K} = \frac{1}{2\lambda}. \tag{7}$$

If $\lambda = 1$, we recover the result of A.1. Evidently $W \to 0$ if $\lambda \to \infty$ (vacancy limit), and $W \to \infty$ if $\lambda \to 0$ (Brownian limit). The above result is confirmed by taking $z \to 0$ in

$$\tilde{r}(z) = 1/[(1-\lambda)z + \lambda\sqrt{(z^2+4)}]. \tag{8}$$

The underlying physical processes seem clear. Suppose the impurity particle is perturbed inelastically. If its mass is very light, it would act as if trapped between two heavy nn atoms which are like walls. The perturbation energy does not become delocalized. If the impurity mass is much heavier than the masses of its neighbors, it begins to resemble a Brownian particle. When inelastically perturbed, it will simply drag the light masses along just as in B, a process we termed ballistic.

### D.1. Periodic diatomic chain

In an infinite chain with periodic boundary conditions we let the masses be either $m_1$ or $m_2$ and arrange them periodically repeated e.g. $m_1 m_2 m_1 m_2 K$. We let $\lambda = m_2/m_1$ now and $A = p_1$, the momentum of anyone of the particles with mass $m_1$. In units $k/m_1 = 1$, the canonical form is given by [10]

$$W = \frac{1 \cdot \lambda \cdot 1 \cdot \lambda \cdot 1 \cdot K}{2\lambda \cdot 1 \cdot \lambda \cdot 1 \cdot K} \tag{10}$$

We note that if $\lambda = 1$, it reduces to that of a monatomic chain given in A.1. The above infinite product is too intricate to be simply determined and we turn to $\tilde{r}(z)$, which is given by

$$1/\tilde{r}(z) = z + 4\lambda(z^2 + \lambda + 1)/[z(z^2+2) + F^{1/2}] \tag{11}$$

where

$$F = (z^2+2)(z^2+2\lambda)(z^2+2\lambda+2). \tag{12}$$

If we set $z = 0$, we obtain a remarkable result

$$W = 1/\sqrt{2\lambda)(\lambda+1)}. \tag{13}$$

If $\lambda = 1$, clearly we recover the monatomic result. If $\lambda \to \infty$, we reach the localization limit, similar to the vacancy limit in the one-impurity mass problem in C. If $\lambda \to 0$, we reach the ballistic limit. In between the model is ergodic with respect to $p_1$.

*D.2. Aperiodic diatomic chains*

Our final example is an assembly of *nn* coupled *HOs* of two different masses $m_1$ and $m_2$ but arranged in a Fibonacci sequence. We form this sequence on a chain of lattice sites as follows: The sites are labeled $-N, \ldots -2, -1, 0, 1, 2, \ldots N$, forming a ring by making $-N$ and $N$ the same site and $N \to \infty$. Put $m_1$ at site 0, $m_2$ on site 1 in a sequence arranged as: Denoting only the subscripts of the masses, 1, 2, 1, 1, 2, 1, 2, 1, 1..., from site 0 to $N$. Also arranged is its reflection about site 0 on sites from $-1$ to $-N$. Thus it is a ring composed of two infinitely long Fibonacci chains, fused at sites 0, and $N$ and $-N$. Each *nn* is coupled harmonically by coupling constant $k$.

We choose $A = p_0$, the momentum of the oscillator at site 0 with mass $m_1$. Let $\lambda = m_2/m_1$ as in the periodic diatomic chain. See D.1. The canonical form of $W$ shows the following structure:

$$W = \frac{x \cdot y \cdot x \cdot y \cdot \mathsf{K}}{2y \cdot x' \cdot y \cdot x' \cdot y \cdot \mathsf{K}} \tag{14}$$

where $x = (1 \cdot \lambda \cdot \lambda \cdot 1)$, $x' = (1 \cdot \lambda \cdot 1 \cdot \lambda)$ and $y = (\lambda \cdot 1 \cdot \lambda \cdot \lambda)$. Note that if $\lambda = 1$, $W$ becomes that for a monotonic chain of A.1. Observe that the aperiodic character has been removed in $W$. Its structure is similar to that of the periodic chain. Hence one is tempted to draw the solution from the periodic case. Since $x, x'$ and $y$ are products of finite sequence, we replace them with their values $\lambda^2, \lambda^2$, and $\lambda^3$, respectively. Taking $x' = x$, divide the infinite products by $x$ above and below repeatedly, and obtain infinite products of $y/x = \lambda$. Now it is exactly in the same form as (10). Hence $W$ is also given by (13). Evidently ergodicity is unaffected when a periodic diatomic chain is changed into an aperiodic one.

## IV. Concluding remarks

The main motivation of our study of ergodicity is to obtain a physical basis for understanding why the ergodic hypothesis works and why it may fail. By taking on one of the simplest many-body models, we have demonstrated the underlying physical mechanisms that enter into the hypothesis. When a particle in a Hermitian model is perturbed, it is a question of what happens to the perturbation energy. If it becomes delocalized, indicated by the existence of a zero frequency mode, one may say that the model is ergodic with respect to the variable representing the perturbed particle. Evidently there may be several situations where the perturbation energy does not become delocalized, indicated by the absence of a zero frequency mode. Then the ergodic hypothesis fails. In this work we have illustrated these properties explicitly through the assemblies of *nn* coupled *HOs*.

## Acknowledgment

The author thanks Prof. D.C. Mattis for drawing his attention to the work of D.W. Jepsen.